\documentclass{moriond}
\usepackage[symbol]{footmisc}

\def\be{\begin{equation}}
\def\ee{\end{equation}}
\def\bea{\begin{eqnarray}}
\def\eea{\end{eqnarray}}

\newcommand{\Photo}{}

\begin{document}
\vspace*{4cm}
\title{SEMILEPTONIC B DECAYS MATRIX ELEMENTS}

\author{$\textsc{G.MARTINELLI}^a$,  $\textsc{M. NAVIGLIO}^{b,c}$\ \footnote[2]{speaker}, $\textsc{S. SIMULA}^d$, $\textsc{L.VITTORIO}^{e,c}$}
\address{	$^a$ Physics Department and INFN Sezione di Roma La Sapienza, \\ Piazzale Aldo Moro 5, 00185 Roma, Italy\\
$^b$Dipartimento di Fisica dell’Universit\`a di Pisa,\\
Largo Pontecorvo 3, I-56127 Pisa, Italy\\$^c$ Istituto Nazionale di Fisica Nucleare, Sezione di Pisa,\\ Largo Bruno Pontecorvo 3, I-56127 Pisa, Italy \\ $^d$ Istituto Nazionale di Fisica Nucleare. Sezione di Roma Tre,\\ Via della Vasca Navale 84, I-00146 Rome, Italy \\ $^e$ Scuola Normale Superiore,
Piazza dei Cavalieri 7, 56126 Pisa, Italy}

\maketitle\abstracts{
We present some applications of the unitarity-based Dispersion Matrix (DM) approach to the extraction of the CKM matrix element $|V_{cb}|$ from the experimental data on the exclusive $B_{(s)} \to D_{(s)}^{(*)} \ell \nu_\ell$ decays. The DM method allows to achieve a non-perturbative, model-independent determination of the momentum dependence of the semileptonic form factors. Starting from lattice results available at large values of the 4-momentum transfer and implementing non-perturbative unitarity bounds, the behaviour of the form factors in their whole kinematical range is obtained without introducing any explicit parameterization of their momentum dependence. We firstly illustrate the effectiveness of the method by considering the case of the semileptonic $B \rightarrow \pi$ decay, which is a good benchmark since the kinematic range is large. Then, we focus on the four exclusive semileptonic $B_{(s)} \to D_{(s)}^{(*)} \ell \nu_\ell$ decays and we extract $|V_{cb}|$ from the experimental data for each transition. The average over the four channels is $|V_{cb}| = (41.2 \pm 0.8) \cdot 10^{-3} $. We find, for the first time, an exclusive value which is compatible with the latest inclusive determination at $1\sigma$ level. We address also the issue of Lepton Flavour Universality by computing pure theoretical estimates of the $\tau/\ell$ ratios of the branching fractions for each channel. In the case of a light spectator quark we obtain $R(D^*) =  0.275(8)$ and $R(D) = 0.296(8)$, which are compatible with the corresponding experimental values within $1.3\sigma$. In the case of a strange spectator quark we obtain $\textit{R}(D_s^*) =0.2497(60)$ and $\textit{R}(D_s) = 0.298(5)$. }

\section{The importance of the study of semileptonic B decays}
B decays are very challenging processes from a phenomenological point of view, principally because of two issues. The first one is the $|V_{cb}|$ puzzle, namely the observation of a tension between the exclusive \cite{FLAG} and the inclusive \cite{Bordone} determination of $|V_{cb}|$ at the level of $\simeq$ 2.7 standard deviations. The second one is the discrepancy between the Standard Model predictions and experiments in the determinations of the $\tau/\mu$ ratios of the branching fractions, the so called $R(D^{(*)})$ anomalies, which represent an important test of Lepton Flavour Universality (LFU). According to HFLAV \cite{HFLAV} the discrepancy is of $\simeq 3.08 \sigma$. 

\section{The Dispersive Matrix Method}
Having these two issues, we need to investigate their nature more deeply. In this sense, it is fundamental to improve the precision with which we compute the form factors entering the hadronic matrix elements. Thus, we introduced a new method, the so called Dispersive Matrix (DM) method \cite{Unitarity} based on an existing work \cite{Lellouch}, whose main features we briefly recall here. Let us consider a generic form factor, $f(t)$, entering the hadronic matrix element of a generic $B \rightarrow Y^{(*)}\ell \nu_\ell$ decay, where $Y$ is generic a meson. Once the following quantities are given, namely 
\begin{enumerate}
\item The values of momentum transfer $t_1,...,t_N$ at which the form factor $f$ have been computed (e.g. on the lattice),
 \item the correspondent $f_1,...,f_N$ values of the form factors in that points,
 \item the susceptibility $\chi$ that are computed on the lattice \cite{Unitarity,Martinelli:2021frl},
\end{enumerate}
then the properties of the method allow us to find bounds on the value of the form factor at a generic value of the momentum transfer. In particular, by defining $z_1, ..., z_N$ where $
z(t) = \frac{\sqrt{\frac{t_+-t}{t_+-t_-}}-1}{\sqrt{\frac{t_+-t}{t_+-t_-}}+1}$ with $t_{\pm} = (m_B \pm m_Y)^2$, the form factor in the point $z$ is bounded by unitarity, analyticity and crossing symmetry to be inside the interval
\begin{equation}
\beta(z) - \sqrt{\gamma(z)}\leq f(z) \leq  \beta(z) + \sqrt{\gamma(z)}
\end{equation}
where 
\begin{equation}
\beta(z) \equiv \frac{1}{\phi(z)d(z)}\sum_{j=1}^N \phi_j f_j d_j\frac{1-z^2_j}{z-z_j},  \ \ \ \ \ 
\gamma(z) \equiv \frac{1}{1-z^2}\frac{1}{\phi^2(z)d^2(z)}(\chi-\chi_{DM}),
\end{equation}
\begin{equation}
\chi_{DM} = \sum_{i,j=1}^N\phi_if_i\phi_jf_jd_id_j\frac{(1-z_i^2)(1-z_j^2)}{1-z_iz_j}.
\end{equation}
Here, $d(z)\equiv\prod_{m=1}^N(1-zz_m)/(z-z_m)$, $d_j \equiv\prod_{m\neq j=1}(1-z_jz_m)/(z_j-z_m)$ and the $\phi_j \equiv \phi(z_j)$ are the values of the kinematical function appropriate for the given form factor \cite{Grinstein} containing the contribution of the resonances below the pair production threshold $t_+$. 
\begin{figure}[htb!]
\centering
 \includegraphics[scale=0.40]{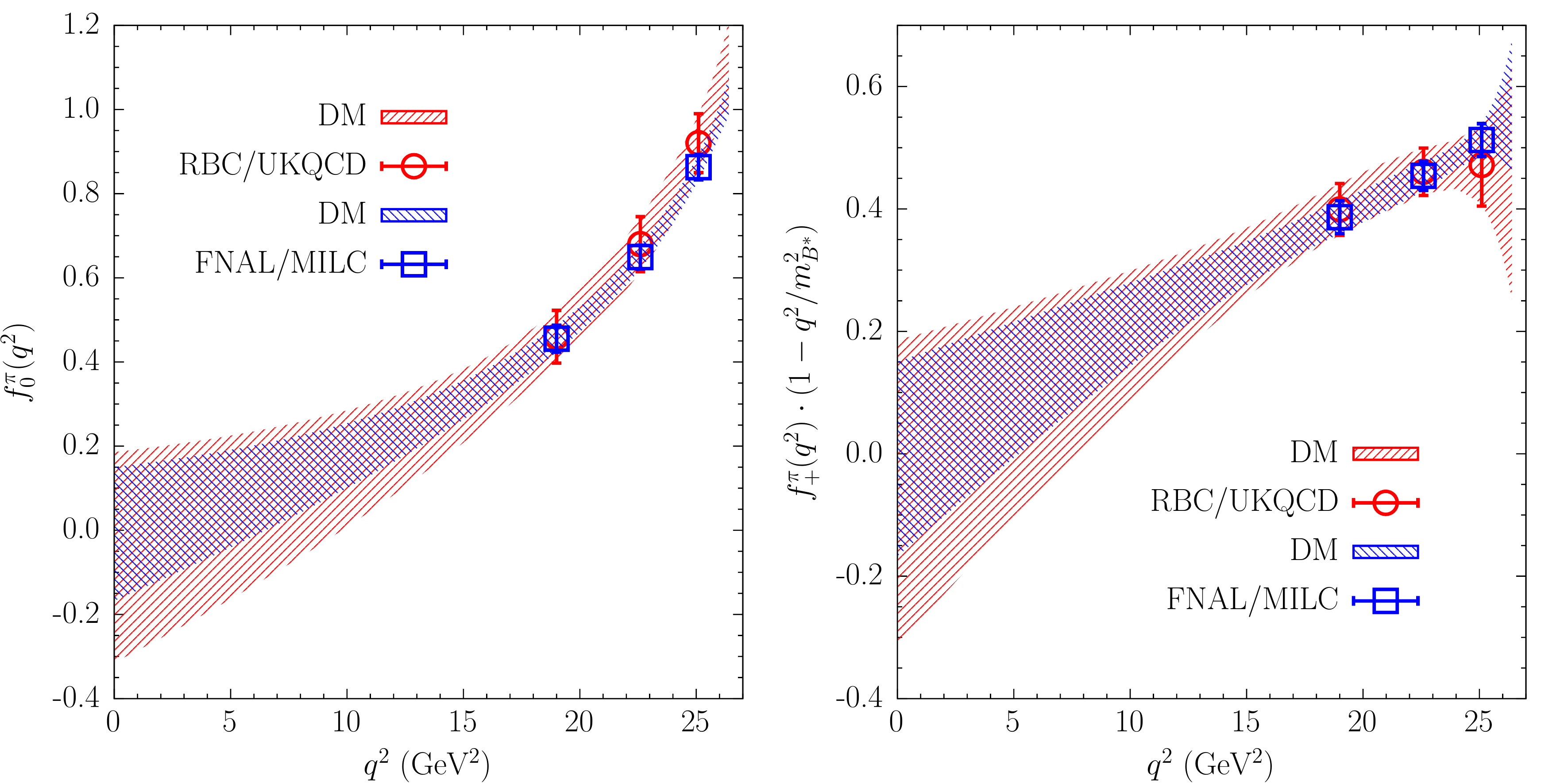}
 \centering
\caption{\textit{The DM bands for the form factors entering the hadronic matrix element of the semileptonic $B\rightarrow \pi$ decay. Note that the only inputs used to build these bands are respectively the red  (RBC/UKQCD) and blue (FNAL/MILC) points.}\hspace*{\fill} \small}
\label{fig1}
\end{figure}
The obtained band of values represents the results of all possible BGL fits satisfying unitarity by construction and passing trough the known points. The results do not rely on any assumption about the functional dependence of the form factors on the momentum transferred. Then, in this sense, they are model independent. Furthermore, the method is entirely based on first principles, the susceptibilities are non perturbative and we do not have series expansions. 
\section{The effectiveness of the method: an illustrative example}
As an example that fully illustrates the effectiveness of the method, we discuss the case of the reconstruction of the form factors entering the semileptonic $B\rightarrow \pi$ matrix element \cite{Pion}. The Figure \ref{fig1} shows the bands, covering the whole kinematic range, obtained using as inputs only the red (RBC/UKQCD) \cite{RBC} and the blue (FNAL/MILC) \cite{FNAL} points. These results show that the DM method allows to make predictions in the whole kinematical range with a quality comparable to the one obtained by the direct calculations, even if only a quite limited number of input lattice data are used. The bands are completely theoretical and come from a non-perturbative and model independent analysis, since no truncated z-expansion are present and no perturbative bounds are used. The method allows us to keep theoretical calculations and experimental data well separated in our analysis, since we do not want to introduce any bias that affects the shape of the form factors.

\section{Main results}
\begin{figure}[htb!]
\centering
 \includegraphics[scale=0.35]{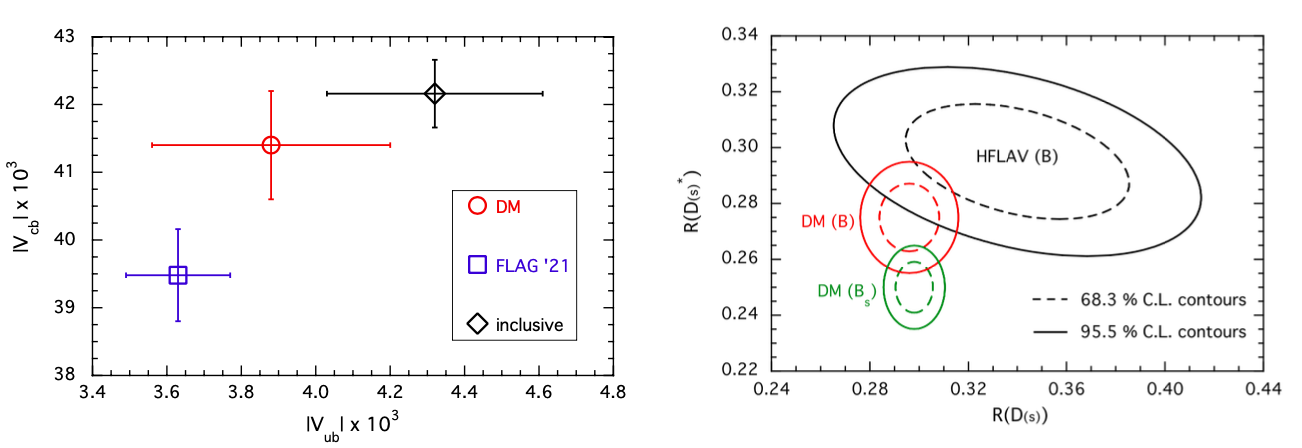}
 \centering
\caption{\textit{The left panel shows the DM values of $|V_{ub}|$ and $|V_{cb}|$ compared with the last results of the FLAG report. The right panel shows the pure theoretical estimates of the ratios $R(D^{(*)})$ compared with the latest averages of HFLAV.}\hspace*{\fill} \small}
\label{fig2}
\end{figure}

At this point we adopted the DM method to analyse the semileptonic $B_{(s)}\rightarrow D_{(s)}^{(*)}$ decays \cite{Martinelli:2021onb,Martinelli:2021myh,Bs_paper}. As in the $B \rightarrow \pi$ case, the method allows us to extract the relevant hadronic FFs in the whole kinematic range using only LQCD results available at large values of the 4-momentum transfer without making any assumption on their momentum dependence. The experimental data are never used to constraint the shape of the FFs but only to extract a determination of $|V_{cb}|$. This allows us also to extract pure theoretical estimates of $R(D)$ and $R(D^*)$. In Table \ref{Vcb} we show the DM results for $|V_{cb}|$ for different channels and the correspondent average. As can be seen, for the first time there is an indication of a sizable reduction of the $|V_{cb}|$ puzzle.
In Table \ref{Observables}, moreover we show our fully theoretical results for LFU and polarization observables. It can be observed also in this case that, for the first time, the $R(D^*)$ anomaly results to be lighter than the $2.5\sigma$ discrepancy stated by HFLAV \cite{HFLAV}. Our findings are graphically collected in Figure \ref{fig2}, where it is also presented the estimate of $|V_{ub}|$ \cite{CKM}.
\begin{table}[t]
\caption[]{DM results for $|V_{cb}|$ for different channels. In the last row we show the corresponding average.}
\label{Vcb}
\vspace{0.4cm}
\begin{center}
\begin{tabular}{|c|c|c|}
\hline
Process & Reference & $~|V_{cb}| \times 10^3~$\\
\hline
Inclusive $b\rightarrow c$  & Bordone et al., arXiv:2107.00604 & $~42.16 \pm 0.50~$\\
\hline
$B \rightarrow D$ & \textbf{DM method} & \textbf{~41.0 $\pm$ 1.2~}\\
& FLAG 2021, arXiv:2111.09849 & $~40.0 \pm 1.0~$\\
\hline
 $B \rightarrow D^*$ & \textbf{DM method} & \textbf{~41.3 $\pm$ 1.7~} \\
 & FLAG 2021, arXiv:2111.09849 & $~39.86\pm 0.88~$ \\
 \hline
 $B_s \rightarrow D_s$ & \textbf{DM method} & \textbf{~41.7 $\pm$ 1.9~ } \\
 \hline
 $B_s \rightarrow D_s^*$ & \textbf{DM method} & \textbf{~40.7 $\pm$ 2.4~} \\
& HPQCD Coll., arXiv:2105.11433 &  $~42.2 \pm 2.3~$\\
\hline 
 \ \  \textbf{Total Mean}    \   &   \ \ \ \ \ \ \ \ \ \ \ \ \  \textbf{DM method} \ \ \ \ \ \ \ \ \ \ \ \ \ &   \ \textbf{~41.2$\pm$ 0.8~ }      \ \\
 \hline
\end{tabular}
\end{center}
\end{table}
\begin{table}[htb!]
\caption[]{Fully theoretical results for LFU and polarization observables. }
\label{Observables}
\vspace{0.4cm}
\begin{center}
{\small
\begin{tabular}{|c|c|c|c|}
\hline
Observable & \textbf{DM method} & Measurements & Difference \\
\hline
$R(D)$ &  \textbf{0.296(8)} & 0.340(27)(13) & $\simeq 1.3\sigma$\\
$R(D_s)$ &  \textbf{0.298(5)} & --- & ---\\
\hline
$R(D^*)$ & \textbf{0.275(8)} & 0.295(11)(8) & $\simeq 1.3 \sigma$\\
$R(D_s^*)$ & \textbf{0.2497(60)}   & --- & ---\\
\hline
$P_\tau(D^*)$ & \textbf{-0.529(7)} & $-0.38(^{+21}_{-16})$ & $<0.3\sigma$ \\
$P_\tau(D_s^*)$ & \textbf{-0.520(12)} & --- & ---\\
\hline
$F_L(D^*)$ & \textbf{0.414(12)} & 0.60(8)(4) & $\simeq 2.0 \sigma$\\
$F_L(D_s^*)$ & \textbf{0.440(16)} & --- & ---\\
\hline
\end{tabular}
}
\end{center}
\renewcommand{\arraystretch}{1.0}
\end{table}

\end{document}